\documentclass[11pt,preprint]{aastex}

\def\iso#1#2{\mbox{${}^{#2}{\rm #1}$}}
\def\be1#1{\iso{Be}{1#1}}
\def\al2#1{\iso{Al}{2#1}}
\def\cl3#1{\iso{Cl}{3#1}}
\def\mn5#1{\iso{Mn}{5#1}}
\def\fe5#1{\iso{Fe}{5#1}}
\def\fe6#1{\iso{Fe}{6#1}}
\def\sm14#1{\iso{Sm}{14#1}}
\def\hf18#1{\iso{Hf}{18#1}}
\def\pu24#1{\iso{Pu}{24#1}}

\def\msol{M_\odot}

\def\avg#1{\langle #1 \rangle}

\def\beq{\begin{equation}}
\def\eeq{\end{equation}}
\def\beqar{\begin{eqnarray}}
\def\eeqar{\end{eqnarray}}

\def\pfrac#1#2{\left( \frac{#1}{#2} \right)}

\def\ga{\mathrel{\mathpalette\fun >}}
\def\fun#1#2{\lower3.6pt\vbox{\baselineskip0pt\lineskip.9pt
  \ialign{$\mathsurround=0pt#1\hfil##\hfil$\crcr#2\crcr\sim\crcr}}}

\begin{document}


\slugcomment{\parbox{2in}{
\begin{flushright}
astro-ph/0410525\\
CERN--PH--TH/2004--190\\
\end{flushright}
}}

\title{Deep-Ocean Crusts as Telescopes: Using
Live Radioisotopes to Probe Supernova Nucleosynthesis}

\author{Brian D. Fields\altaffilmark{1}
\altaffiltext{1}{also Department of Physics, University of Illinois}
}

\affil{Center for Theoretical Astrophysics,
Department of Astronomy, University of Illinois,
Urbana, IL~61801, USA}

\author{Kathrin A. Hochmuth}
\affil{Department of Physics, University of Illinois,
Urbana, IL~61801, USA}

\and

\author{John Ellis}
\affil{Theory Division, Physics Department, CERN, 1211~Geneva~23, 
Switzerland}

\begin{abstract}
Live \fe60 has recently been
detected in a deep-ocean ferromanganese crust,
isolated in layers dating from about 3 Myr ago.
Since \fe60 has a mean life of 2.2~Myr, a near-Earth supernova is 
the only likely 
source for
such a signal, and
we explore here the consequences of a supernova origin.
We combine the \fe60 data with
several supernova nucleosynthesis models to 
calculate the supernova distance as a function of
progenitor mass, finding
an allowed range of $15-120$ pc. 
We also predict the signals expected for
several other radioisotopes, which
are independent of the 
supernova distance.  
Species likely to be present near or above background levels are
\be10, \al26, \mn53, \hf182 and \pu244.
Of these, \hf182 and \pu244 are nearly background-free,
presenting the best opportunities to provide strong confirmation of the
supernova origin of the \fe60 signal, and to
demonstrate that at least some supernovae are the source for
the {\em r}-process.
The accuracies of our predictions are hampered
by large uncertainties in the predicted \fe60
yields for supernovae of different masses, so the new crust data
motivate a redoubled theoretical
attack on this problem.
\end{abstract}


\section{Introduction}

The possibility of finding terrestrial signatures of near-earth 
supernova explosions
has been intensively discussed
over the last decade \citep[see, e.g.,][]{efs,fe}. 
Recent evidence for a supernova remnant
$12^\circ$ in diameter, that may be associated with
\al26 $\gamma$-rays, points to an explosion
within the past few Myr and within
100--200 pc \citep{antlia}.
Moreover, the existence of the Local Bubble -- containing hot,
rarefied gas in which the Sun resides \citep{frisch} -- suggests
that we live in the {\em interior} of one or more supernova
remnants \citep[e.g.,][]{sc}.
This enhances the prospects
of identifying terrestrial evidence of
supernova remnant engulfing the Solar System and the Earth.

Even if it were too far away to be
hazardous for life on Earth
\citep[see, e.g.,][]{ruderman,ES},
a sufficiently
close
supernova explosion 
would deposit on the Earth
supernova ejecta and related material
produced by cosmic rays
\citep{efs}.
This could have resulted in the possible
uptake of supernova-produced elements in deep-ocean sediments or
manganese-iron crusts. Following \citet{efs},  
\citet{knie99}
discovered the first evidence for live \fe60 in a deep-ocean
ferromanganese crust, at a level $\sim 100$ times
the expected background
and consistent with an event in the last
$\sim 5$ Myr at a distance $\sim 30$ pc
\citep{fe}.
However, the structure and geometry of the crust 
made it difficult to reconstruct a reliable
time history, and thus the nature of the \fe60 signal remained uncertain. 

Very recently,
\citet{knie} have studied a different ferromanganese crust,
which is better suited for a high-precision study
with good time resolution. They find a 
\fe60 spike far above background in 
a crust layer which formed between 2.4 and 3.2 Myr ago.
This detection confirms the earlier result,
but now goes beyond detecting the presence of \fe60 to
revealing the details of its deposition history.
The presence of a live radioisotopic signal in a single, 
isolated layer agrees with the straightforward expectations
of the supernova hypothesis \citep{efs,fe}.
These data greatly strengthen the case that supernova
ejecta were deposited on the Earth about 3 Myr ago.  This is a major
new result, and one that finally gives a strong direct empirical basis
for longstanding speculation and arguments based on indirect evidence
regarding the possibility of near-Earth supernova explosions in
the geological past. 

In light of the \citet{knie} result,
the time is ripe to take the supernova
hypothesis seriously and investigate its consequences
quantitatively. Specifically, in this paper we show how ferromanganese 
crusts can be used as telescopes to probe directly the {\em r}-process in 
supernovae.

\section{Summary of the Data}

\citet{knie} measured \fe60 and \mn53 (mean life $\tau_{53} = 5.3$ 
Myr)
throughout the ferromanganese crust 237KD, using
accelerator mass spectrometry techniques.
Over the 28 layers studied, 
just 69 live \fe60 atoms were detected.
This signal was found uniquely in 
layers corresponding to a time interval~\footnote{
The crust was analyzed with
two different time resolutions but
even the smaller timespan of 0.3~Myr in the finer
resolution analysis is expected to be an
artifact representing an upper limit on
the physical deposition timescale,
which is likely to be of order
$\sim 10$~kyr.} 
\beq
t = 2.4-3.2 \ {\rm Myr}
\eeq
before the present.
The resulting isotopic fraction is
\beq
\frac{\fe60}{{\rm Fe}} = (1.9  \pm 0.24) \times 10^{-15},
\eeq
which is a highly significant result, despite its small magnitude,
testifying to the sensitivity and precision of the
experiment.
The principal uncertainty in this result
is due to instrumental background, essentially the contamination due to
stable isobars.

Given the crust size and Fe composition,
the isotopic fraction can be used to 
infer a surface density of \fe60.
To do this, one must also
correct for radioactive decay
and geometric effects,
and the result is
\beq
N_{60,{\rm obs}} = 2.9 \times 10^6  \ {\rm atoms/cm^2}.
\eeq
In order to infer the incident fluence of \fe60,
this result must be corrected for 
geometry and an uptake factor.
Assuming the terrestrial fallout is isotropic,
a geometric factor 1/4 
arises from the ratio of the Earth's full cross section ($\pi 
R_\oplus^2$) 
to its total surface area ($4\pi R_\oplus^2$). 
Also, the crust incorporates only 
a certain percentage of all deposited material incident
upon it, and
a so-called `uptake factor' $U < 1$
accounts for this effect
\citep[for further discussion, see][]{knie,knie99}, leading to
\beq
\label{eq:surf}
N_i(t)=\frac{U}{4}F_i\exp(-t/\tau).
\eeq
The uptake factors are different for each sample and every element,
and thus cannot be evaluated independently of the
details of the sample.
Consequently, in the following discussion
we present results for $U = 1$,
with the understanding that 
all the results presented must be reduced by the appropriate uptake 
factor, which may be inferred from studies of stable isotopes.

\citet{knie} use Mn and Fe data
to infer an iron uptake factor $U_{\rm Fe} \simeq 0.006$,
from which they infer an incident fluence of
\beq
F_{60,\oplus} = 2.0 \times 10^9  \ {\rm atoms/cm^2}.
\eeq
This signal encodes key information about the supernova
explosion, as we now see.

\section{Predictions:  Supernova Distance and Other Radioisotope Signals}

The deposited amount of supernova-produced material of an isotope $i$ is
dependent on the distance $D$ of the event and the total ejected mass
$M_i$. The fluence, i.e., the number of atoms per unit area that arrive at 
the Earth, is \citep{fe,efs}:
\beq
\label{eq:flu}
F_i=\frac{M_{{\rm ej}, i}}{4\pi A_i m_p D^2 }, 
\eeq
where $A_i$ is the mass number of isotope $i$ and $m_p$ is the proton mass.
Present-day measurements of the fluence must also
be corrected for decay. 

The isotopic fraction of $i$ in the crust is determined
by the relative amounts of supernova ejecta and
substrate deposited within the observed time interval.
If the substrate material has density $\rho$
and deposition rate $\dot{h}$,
then the total mass flux onto the crust
is $j = \rho \dot{h}$,
and the iron flux onto the crust
is $X_{\rm Fe} j$, where
$X_{\rm Fe} \sim 0.2$ 
\citep{hein}
is the iron mass fraction in 
the crust, and other 
elements $i$ can be treated in a corresponding way.
Finally,
if the supernova signal is measured with a time
resolution $\Delta t$, the
surface density of crust atoms deposited is
$N = X_{\rm Fe} j \Delta t/A m_p$, where $A \approx A_i$ is
the mean mass number of the element, and so the
isotopic fraction of $i$ in the crust is
\beqar
\label{eq:isofrac}
\frac{N_i}{N} 
  & = & \frac{U_i M_{{\rm ej},i}}{16 \pi X_{\rm Fe} \rho \dot{h} \Delta t D^2} 
       e^{-t/\tau_i} \\
  & = & 1.9 \times 10^{-15} \
      \pfrac{M_{{\rm ej},i}}{2 \times 10^{-5} \msol} \
      \pfrac{0.3 \ {\rm Myr}}{\Delta t} \
      \pfrac{30 \ {\rm pc}}{D}^2 ,
\eeqar
where the fiducial numbers are for the \fe60
signal at 3 Myr,
with $\rho = 5 \ {\rm g \, cm^{-3}}$
and $\dot{h} = 2.5 \ {\rm mm \ Myr^{-1}}$.

The confirmed observation of a live \fe60 signal
spurs us to predict the inventory of all
long-lived radioisotopes.
For each, given the yield $M_{{\rm ej},j}$,
one can readily compute the fluence and/or
surface density, based on the \fe60 signal:
\beq
N_j = \frac{U_j}{U_{\rm Fe}} \frac{60} {A_j}
  \frac{M_{{\rm ej},j}}{M_{\rm ej,60}} N_{60}
  e^{(1/\tau_{60}-1/\tau_i)t},
\eeq
a result that is independent of
supernova distance $D$.
Consequently, detection 
of any radioisotopic species $j$ would immediately
measure the ratio of its yield relative to
\fe60, and thus probe supernova nucleosynthesis -
assuming that the uptake factors $U$ are known, and that the Earth receives
an isotropic `fair sample' of all supernova ejecta~\footnote{The uptake 
factors $U$ can be obtained from studies of stable isotopes or, in the 
case of \pu244, from studies of the recent deposition of anthropogenic Pu 
on the surface of the crust~\citep{wallner}.}.
Indeed, even a significant upper limit could provide
important information.

The explosion signals divide into radioisotopes from
three sources:
supernovae, the
{\em r}-process and cosmic rays, and we consider each of these in turn.

\subsection{Supernova Isotopes}

To get an impression of the distance of the supernova, one needs 
to know the \fe60 yield $M_{\small \fe60}$ of a 
supernova. This and the yields of other isotopes have been calculated 
in \citet{ww},
and more recently again in \citet{rhhw}. 
The 12 supernova explosion models of \citet{ww} 
span a mass range from $11-40M_\odot$, and
the 5 newer calculations in \citet{rhhw}
cover the range $15-25M_\odot$. 
The latter give a higher \fe60 yield than those calculated 
in \citet{ww}. 
A common feature of the \fe60 yields in the two papers 
is strong variation over the observed mass range, which
attests to the difficulty of calculating the production of
this element, reflected also in differences between the results of 
\citet{rhhw} and \citet{ww} for 
supernovae of similar masses. 
Nevertheless, we use these mass yields together with eq.~(\ref{eq:flu}) 
to calculate the possible supernova distance $D$, albeit with 
considerable uncertainties induced by the different mass yields.

\begin{figure}
\includegraphics[width=0.6\textwidth,angle=270]{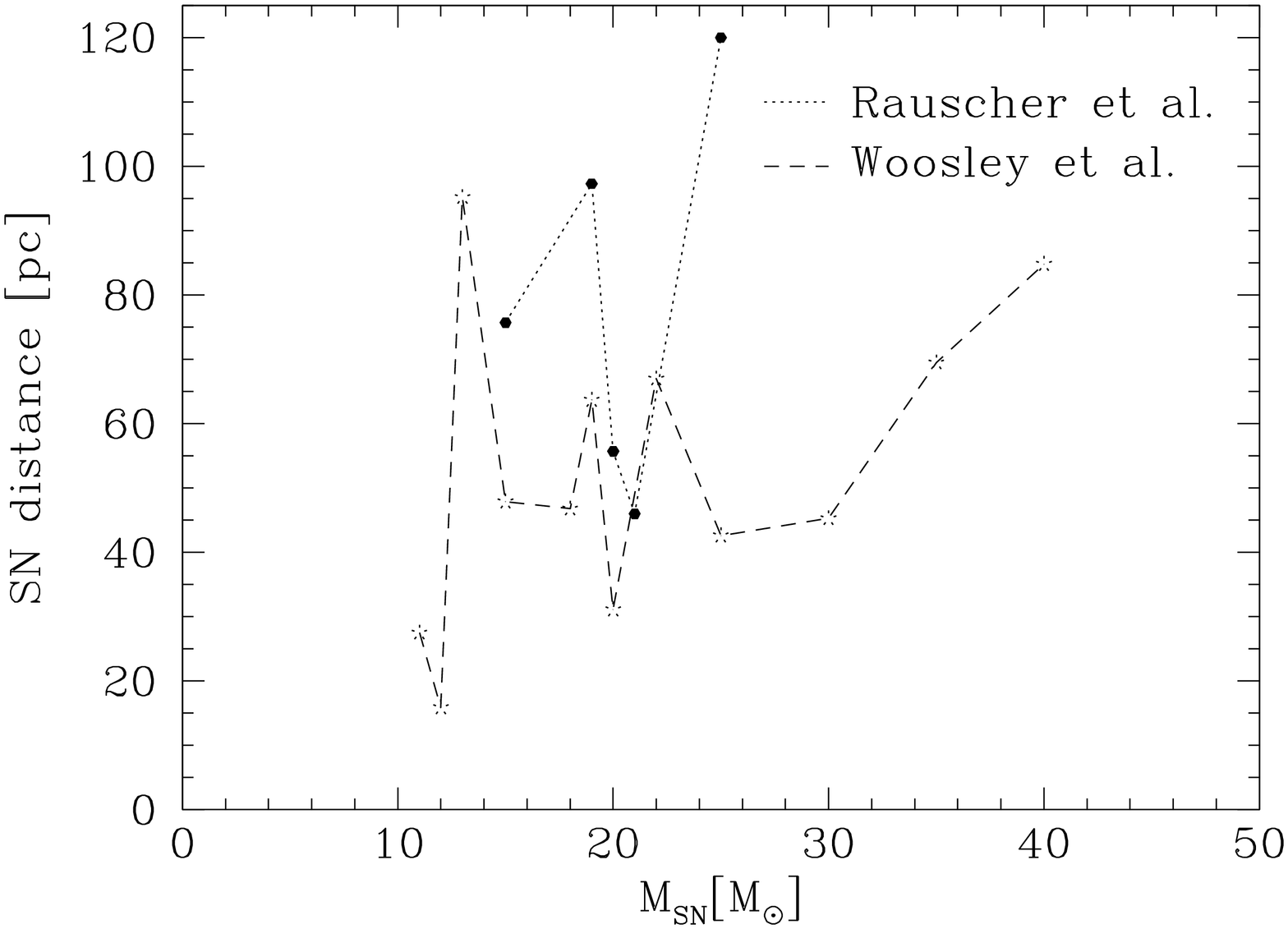}
\caption{\it
The supernova distance implied by the \citet{knie}
result, shown as a function of progenitor mass.
Results are presented for the theoretical yields of
\citet{rhhw} and \citet{ww}.
The curves scale with the \fe60 yield as $M_{{\rm ej}, 60}^{-1/2}$
(\ref{eq:isofrac}), and the
large variations in distance 
reflect the large variations in the yields with
progenitor mass, and the differences between the curves
give a sense of the range of theoretical uncertainty in the yields.
}
\end{figure}

The results of this calculation can be seen in Fig.~1, where
we see that the distances cover a range between $\sim 10$ pc and $\sim 
100$pc. 
This range is similar to that found in the
similar analysis of \citet{knie}.
More generally, this distance range
falls within limits set by 
\citet{ES} 
and \citet{efs}, 
who note that a supernova closer than 10 pc could be 
devastating to
life on earth \citep[e.g.,][]{ruderman}, 
whereas a supernova that was too far away would not have 
the power to overcome the solar wind and 
deposit a noticeable amount of isotopes. 

Having estimated the distance, it is now easy to obtain the surface 
densities
(\ref{eq:surf}) of other sufficiently long-lived supernova-produced
elements like \cl36 ($\tau=0.435$ Myr),
\mn53 ($\tau=4.5$ Myr) and \al26
($\tau=1.03$ Myr), using again the values of \citet{rhhw} and \citet{ww} 
for the ejected masses of
\cl36, \mn53 and \al26. The results for \mn53 and
\al26 are shown in Figure 2, and vary wildly, due to
the variations in the \fe60 mass yields. 
A better prediction of the
\fe60 yields would therefore be a valuable input into future calculations. 
On the other hand, the detection in the sample of \citet{knie} (or in
comparable samples) of other
isotopes, besides \fe60, whose yields are not as wildly varying
(e.g. \al26), would allow for a more accurate calculation of the
distance and hence stronger predictions for the surface
densities of other supernova isotopes,
and possibly the supernova mass as well. 
At the present stage, the
results shown in Figure 2 can be used only to get a feel
for the expected order of magnitude (without uptake factors) of the
surface densities of supernova isotopes. As we see in the next
Section, the corresponding results for \cl36 are not worth showing
because of the high expected background.

\begin{figure}
\includegraphics[width=0.45\textwidth]{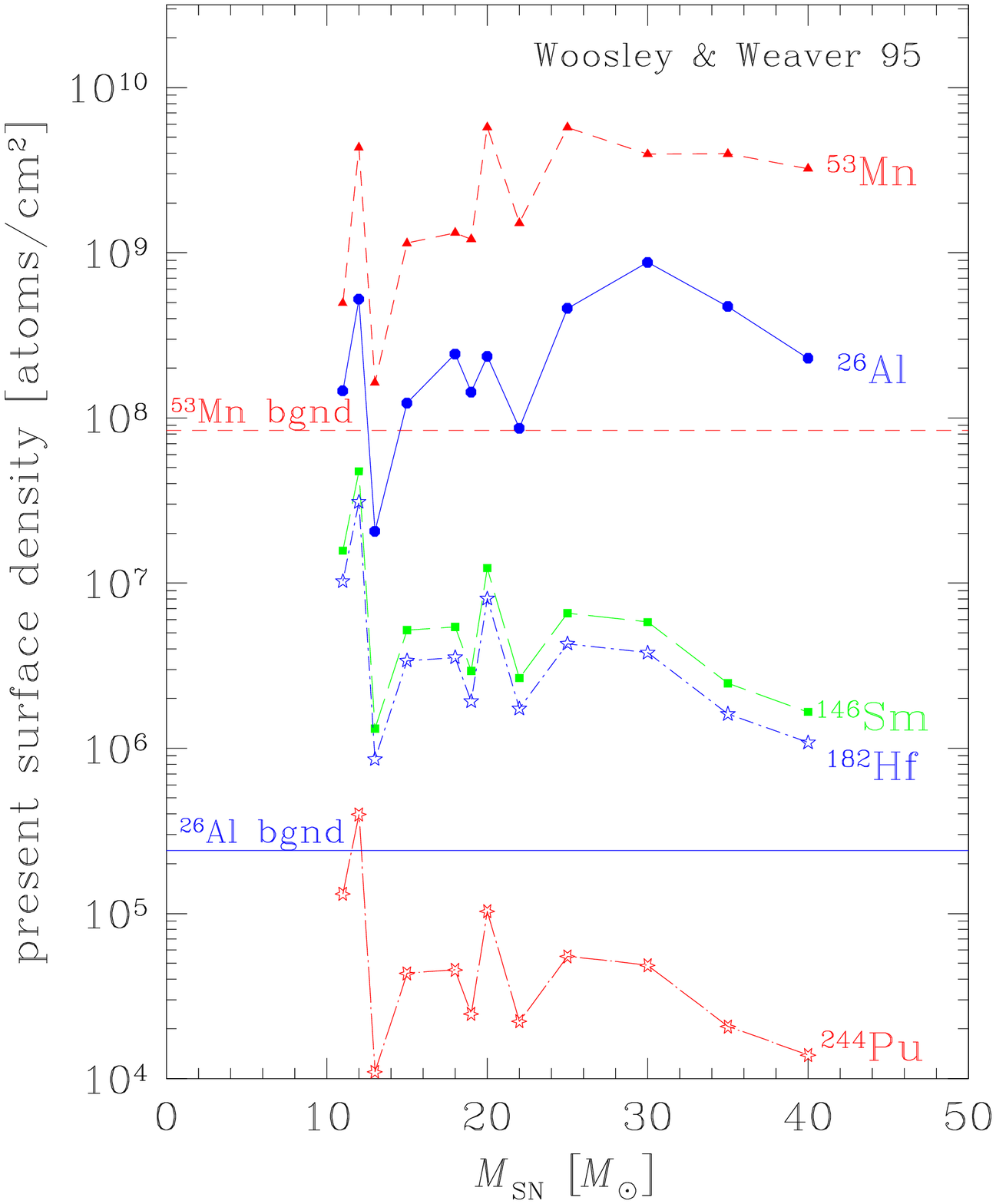}
\includegraphics[width=0.45\textwidth]{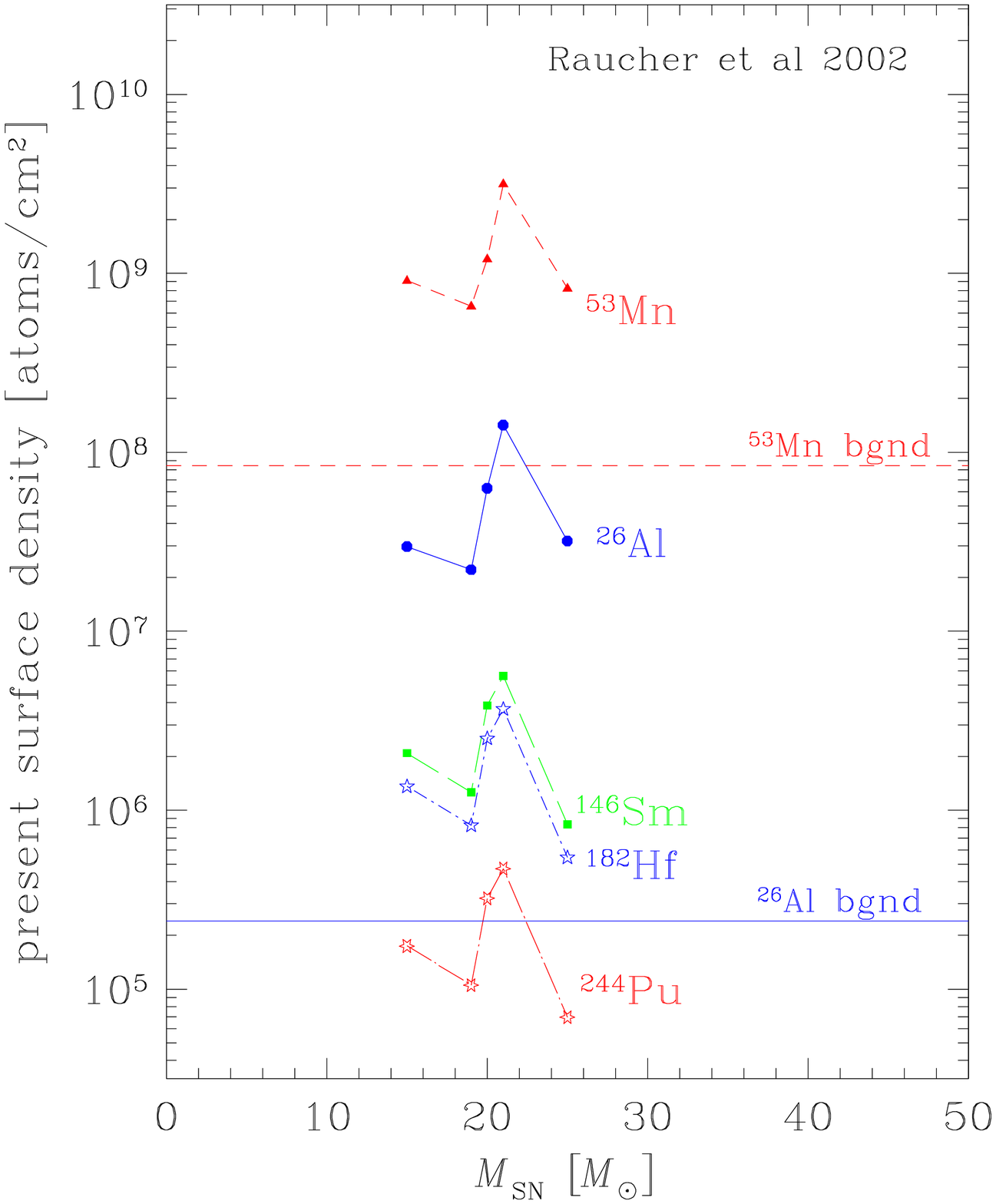}
\caption{\it
Predicted terrestrial surface densities $N_i$ as functions of
supernova progenitor mass (eq.~\ref{eq:flu}).
For \mn53 and \al26, the background levels are indicated.
Results are corrected for geometry and decay,
but are not corrected for 
site- and element-dependent
uptake factors $U$
which appear in eq.~(\ref{eq:surf}).
As in Figure 1, 
variations in the predicted \fe60 yields
are the dominant cause of the large excursions with varying mass.
{\em (a)} Results using \citet{ww} yields.
{\em (b)} Results using the \citet{rhhw} yields.
}
\end{figure}

\subsection{{\em r}-Process Elements}

The {\em r}-process elements present
a special case.
These are nuclides which are created when
`seed' nuclei -- typically iron-peak elements -- are
exposed to a rapid burst of neutrons.
Whilst the microphysics 
and thermodynamical requirements for the {\em r}-process 
are fairly well understood,
its astrophysical site is not yet known;
both core-collapse supernovae \citep[e.g.,][]{wwmhm}
and neutron-star mergers \citep{elps} 
have been studied in some detail.
Consequently, a firm 
correlated detection of a known supernova-produced isotope 
(e.g., \cl36, \mn53, \al26, or
\fe60) with an {\em r}-process isotope
(\pu244, \sm146, or \hf182) 
would be of the utmost importance: it
would provide direct proof that at least some supernovae 
do generate {\em r}-process elements. 
Intriguingly, \citet{wallner}
report the detection of a single live \pu244 atom
in the same crust that contained the original
live \fe60 signal \citep{knie99}.
One obviously should not overinterpret such a marginal result,
but we are encouraged that \pu244 searches are becoming
feasible, and we advocate strongly further searches
for all {\em r}-process elements.

In order to get a feel for the expected order of magnitude of the
surface density of \pu244, \sm146 and
\hf182, we must determine the supernova yields of these
elements. We do this empirically, by noting that,
in metal-poor Galactic halo stars, the heaviest
elements are always distributed in the same ratios 
as the solar {\em r}-process component.  This
result implies that, at least for the heaviest elements
(including the species of interest here), the
{\em r}-process yield distribution is `universal', i.e.,
always produced in the solar ratios: \citep{ftc} and 
refs.~therein.
Thus, all {\em r}-process isotopes are fixed once one is known.
We make this normalization using the halo star observations,
and choose Eu.
\citet{ftc} use observations of Eu and Fe in
halo stars to deduce that
one supernova produces 
$m_{\rm Eu}
   \approx 5 \times 10^{-7}M_\odot$.
It then follows that any other species $r$
has 
$m_r=({A_r}/{152})({r}/{\rm Eu})_{\odot,r} m_{\rm Eu}$.
For solar abundances of the {\em r}-process elements, 
we use a recent theoretical model
\citep{pfeiffer,cowan}, 
from which we extract the 
following ratios:
$({\rm Sm/Eu})_{\odot,r}=1.1$,
$({\rm Hf/Eu})_{\odot,r}=1.1$,
and $({\rm Pu/Eu})_{\odot,r}=0.12$;
this gives
supernova yields 
\beq
m_{\rm Sm}=5.4 \times 10^{-7}M_\odot, \ \
m_{\rm Hf}=6.8 \times 10^{-7}M_\odot, \ \
m_{\rm Pu}=0.88 \times 10^{-7}M_\odot.
\eeq

Using eq.~(\ref{eq:surf}), one can now calculate the surface densities of
the {\em r}-process elements in question, and our results are shown in 
Figure 2.
Notice again the wild variations at the present stage, which
allow the shown results only to give an impression of the expected 
orders of
magnitude of the surface densities.  

Estimates of the expected backgrounds are given in the next Section.

\subsection{Cosmic-Ray Production: $\stackrel{10}{}$Be}

Finally, we consider the possibility of a supernova signal due to 
$\stackrel{10}{}$Be by cosmic rays.
The explosion of a nearby supernova enhances the cosmic ray flux at
Earth, and the element that is dominantly produced is $\stackrel{10}{}$Be
($\tau=2.18$ Myr).
\citet{efs} estimate that 
an average supernova will put a
fraction $\xi_{\rm SN} \ga 1/100$ of its mechanical energy 
$E_{\rm SN}$ 
into the production of cosmic rays with an average energy of 
$\avg{E_{\rm CR}}=1$ GeV. Thus we get a cosmic ray fluence of
\beq
\label{eq:crflu}
\Phi \Delta t=\xi_{CR}f_{CR}\frac{\avg{E_{SN}}}{4\pi D^2 \avg{E_{CR}}},
\eeq
The deflecting effects of the solar wind
mean that only some fraction
$f_{CR}$ of cosmic rays penetrate to the Earth.
Following \citet{fe}, we take $f_{CR} = 1/10$.

The enhanced cosmic-ray fluence at Earth
increases radioisotope production in the atmosphere,
according to
\beq
\label{eq:branch}
F_i=Y_i\Phi_p\Delta t,
\eeq
where $Y_i$ is the branching ratio for cosmic-ray production of 
isotope $i$; a useful tabulation of these appears in
\citet{obrien}. Using eq.~(\ref{eq:branch}) with 
$Y(\be10)=1.36 \times 10^{-2}$,
$\Phi$ from eq.~(\ref{eq:crflu}),
and the distances from Figure 1,
one finds a $\stackrel{10}{}$Be fluence of at most
$2.8 \times 10^8 {\rm at/cm^2}$, 
and a surface density of at most $2.0 \times 10^7 {\rm atoms \, cm^{-2}}$.

\section{Backgrounds}

We now consider the dominant sources of backgrounds to these various
signals, which originate from different sources.

\subsection{Background for \mn53 and \al26: Meteoritic Infall}

The backgrounds for \al26 and \mn53 are dominated by the infall of
meteorites. These have been exposed to cosmic rays, causing an enhancement
of certain elements due to spallation. An introductory discussion of
meteoritic backgrounds can be found in \citet{fe}. The flux $\Phi_i$ of
infalling isotopes can be calculated as
\beq
\label{eq:metr}
\Phi_i=\frac{JX_i}{4\pi R_\oplus^2 A_i m_p},
\eeq
where $J\approx 4 \times 10^{10}{} g/yr$ is the infalling meteoritic mass 
rate \citep{pe,et,lb} and 
$X_i$ is the mass fraction of isotope $i$ found in Solar-System 
meteorites.
\citet{michel} examine the production of isotopes in meteorites due to
cosmic rays. They give their main tabulated results in terms of the
specific activity $\Gamma_i$, which is the number of decays per minute
and kg of iron or rock, respectively, for \mn53 or \al26. We take the
fiducial values of 400 dpm/kg Fe and 40 dpm/kg Si. In fact, \al26 has
more than one production branch, but as Si is the most dominant
element we will use it for an estimate of the background.  We assume
the infalling meteorites to have an iron mass fraction $X_{Fe} = 0.19$ and 
a silicon mass fraction $X_{Si} = 0.25$ \citep{ag}. 
For a meteorite with such mass
fractions, we obtain $X_i=m_i\tau_i\Gamma_iX_{Fe/Si}$ and therefore
$X(\mn53) = 1.9 \times 10^{-11}$ and
$X(\al26) = 2.4 \times 10^{-13}$.
With these, we find fluxes of 
\beqar
\Phi(\mn53) & = & 1.7  \times 10^{9} \ {\rm Myr^{-1}cm^{-2}}, \\
\Phi(\al26) & = & 4.3  \times 10^{7} \ {\rm Myr^{-1}cm^{-2}}.
\eeqar
In order to calculate the background fluence and surface density 
we have to multiply the flux by the time resolution.
Recalling that the \fe60 signal in \citet{knie} spreads over a time of 
approximately $\Delta t \simeq 1/3$ Myr,
we obtain surface densities of
\beqar
N(\mn53) & = & 8.4 \times 10^7 {\rm atoms \, cm^{-2}}, \\
N(\al26) & = & 2.4 \times 10^5 {\rm atoms \, cm^{-2}}.
\eeqar
In comparison with the signal strengths of order 
$10^8-10^9{}~ {\rm atoms \, cm^{-2}}$ for \mn53
and $10^7-10^8{} ~ {\rm atoms \, cm^{-2}}$ for
\al26, these backgrounds are negligible.

\subsection{Background for \be10 and \cl36:  Cosmic-Ray Interactions}

A different approach has to be taken for calculating the backgrounds to 
\cl36
and \be10,
as these nuclides are primarily produced by the 
direct interaction of cosmic rays with the atmosphere. 
Using eq.~(\ref{eq:branch}),
with the ambient cosmic-ray proton flux 
$\Phi_p \simeq 10{} \ {\rm cm^{-2}s^{-1}}$ 
and a time resolution $\Delta t=0.3$ Myr,
for \cl36 one finds $Y(\cl36)=4.74  \times 10^{-4}$.
Consequently, one obtains a present-day surface density of
\beq
N(\cl36)=2.0  \times 10^7 {\rm atoms \, cm^{-2}}.
\eeq
This is to be compared with a signal strength $\sim 10^4-10^6{} {\rm atoms
\, cm^{-2}}$ and we therefore do not expect the detection of \cl36 in the
ferromanganese crust sample of \citet{knie} or in similar samples.

\subsection{Background for \pu244, \sm146, and \hf182:  Fission} 

As \sm146, \hf182 and \pu244 are extinct radioactive isotopes, they do not
have a significant abundance in Solar-System meteorites. Moreover, they
are not produced in sufficient amounts by cosmic ray spallation. 
The only background source is spontaneous fission of actinides.
Such decays lead to a distribution of daughter products,
but these only contribute significantly to \sm146,
not to \hf182.  

We estimate the spontaneous fission background of
\sm146 as follows.  The dominant parent nucleus
is \iso{U}{238}, with 
a fission lifetime $\tau_{238,{\rm sf}} = 1.2 \times 10^{16}$ yr.
If the fission branching ratio to \sm146 is
$f \sim 0.1$, then the equilibrium \sm146 fission background
thus has $(\sm146/\iso{U}{238})_{\rm eq} = f \tau_{146}/\tau_{238,{\rm sf}}$.
The \sm146 background flux onto the crust is then
$\Phi_{146} = (\sm146/\iso{U}{238})_{\rm eq} X_{238} j/238 m_p$
(where $j$ is the total mass flux onto the sample, 
and $X_{238} \sim 3 \times 10^{-6}$ is the mass fraction
of \iso{U}{238} in the Earth's crust).
Thus the background accumulated in a resolution time
$\Delta t \sim 0.3$ Myr is
\beq
N_{146,{\rm bgd}} 
  = U f \frac{\tau_{146}}{\tau_{238,{\rm sf}}} X_{238}
    \frac{j \Delta t}{238 m_p}
  \simeq 3.6 \times 10^7 U
\eeq
which is larger then our prediction by about an order of magnitude.
Thus we conclude that the \sm146 is probably too small to see.

Since \pu244 is {\em heavier} than ambient U and Th, it is not produced by
fission. Apart from recent human activity, the only known possible source
of \pu244 is the {\em r}-process. Moreover, anthropogenic \pu244 is
expected to be negligible in the ferromanganese crusts of interest.

\section{Discussion}

Following the prediction of \citet{efs}, \citet{knie99} found initial
evidence for supernova \fe60 in a ferromanganese crust, prompting
\citet{fe} to discuss other isotopic signatures of a nearby supernova. The
exciting new result of \citet{knie} confirms the earlier observation with
higher significance and more precise dating, motivating this paper. Using
the data of \citet{knie} to estimate the local interstellar fluence of
\fe60 and combining this with the total mass of \fe60 ejected by diverse
supernovae \citep[][]{rhhw,ww}, 
we have inferred a supernova distance range 
$D \sim 10-100$ pc,
consistent with \citet{knie}.

The \fe60 detection and nucleosynthesis models
together allow us to make distance-independent
predictions for other
radioisotopes.
For supernova-produced isotopes such as \al26, \mn53 and \cl36,
we find that yields of the
first two of these isotopes are likely to be detectable above background.
We also consider the {\em r}-process elements \pu244, 
\hf182, and \sm146, finding again that the first two
are likely to exceed background. 
If the supernova responsible for the \fe60 observations of
\citet{knie} was also a site for the {\em r}-process, clear signals should
be visible. This would finally provide direct proof that supernovae are
sites for the {\em r}-process.

The firm detection of \fe60 also adds support to
recent ideas regarding the origin of the 
Local Bubble.
\citet{sc} concluded that several events are needed
to create the Local Bubble, the last occurring
within the past 5 Myr in order to maintain the 
high interstellar temperature.
\citet{ma} argued that the need for multiple events
points to correlated explosions within an OB
association, and specifically suggested
the Lower Centarus Crux (LCC) of the Sco-Cen association as
the likely site.
\citet{bmc} have examined the possibility that this OB
association could have been the site of the \fe60 progenitor.
From kinematic considerations they find that the
closest approach of the LCC was about
100 pc, with considerable uncertainties.  
Taken at face value, this result is in the upper part
of our range of distance estimates.
This suggests that either the event was a distant
one (and thus the \fe60 yield was high), or that
the LCC approach was closer than estimated, or
that the LCC was not the event site.

The accuracies of our predictions for the fluences of these isotopes are
limited by the wide variations in the \fe60 yields found in calculations
for different progenitor masses, and between different model calculations
\citep[][]{rhhw,ww}. Therefore, at the present stage our results give only
approximate estimates of the expected orders of magnitude of the fluences
of supernova isotopes. This underlines the great need for more theoretical
studies of \fe60 production in supernovae, able to reduce the
calculational uncertainties. Such a reduction would be necessary before 
one could predict any biological effects that might be associated with the 
supernova responsible for the \fe60 observations of \citet{knie99} and 
\citet{knie}.

Measurements using other ferromanganese crusts would be most welcome, 
and observations of other supernova radioisotopes have high priority. 
These would enable one to evade the uncertainties associated with the 
\fe60 yield.
Another helpful calibration point 
comes from the recent evidence \citep{rhessi}
for the $\gamma$-ray line signal from
the decay of interstellar \fe60.
This reflects the current Galactic \fe60 inventory, and
constrains \fe60 nucleosynthesis
averaged over all sources
\citep{prantzos}.

The important results of \citet{knie99} and \citet{knie} open up a new era
of `deep-ocean crust astronomy'. After providing already the first
observations of the terrestrial impact of a nearby supernova, their
measurements now offer the prospect of verifying directly that the {\em
r}-process takes place, at least partly, in ejecta from supernovae.

\acknowledgments
We thank G\"{u}nther Korschinek and Klaus Knie for
invaluable discussions regarding their \fe60 data.
This material is based upon work supported by the National Science
Foundation under Grant No. AST-0092939

\end{document}